\documentclass[10pt,letterpaper,twocolumn]{article} 

\usepackage{ol2}
\usepackage[draft,implicit=false]{hyperref}
\usepackage{amsmath}

\begin{document}

\twocolumn[ 

\title{Observation of soliton explosions in a passively mode-locked fiber laser}

\author{Antoine F. J. Runge,$^{1,*}$ Neil G. R. Broderick,$^{1,2}$ and Miro Erkintalo$^{1}$}

\address{
$^1$Department of Physics, The University of Auckland, Private Bag 92019, Auckland 1142, New Zealand\\
$^2$Dodd-Walls Centre, The University of Auckland, Private Bag 92019, Auckland 1142, New Zealand
$^*$Corresponding author: arun928@aucklanduni.ac.nz
}

\begin{abstract}
Soliton explosions are among the most exotic dissipative phenomena studied in mode-locked lasers. In this regime, a dissipative soliton circulating in the laser cavity
experiences an abrupt structural collapse, but within a few roundtrips returns to its original quasi-stable state. In this work we report on the first observation of such events in a fiber laser. Specifically, we identify clear explosion signatures in measurements of shot-to-shot spectra of an Yb-doped mode-locked fiber laser that is operating in a transition regime between stable and noise-like emission. The comparatively long, all-normal-dispersion cavity used in our experiments also permits direct time-domain measurements, and we show that the explosions manifest themselves as abrupt temporal shifts in the output pulse train. Our experimental results are in good agreement with realistic numerical simulations based on an iterative cavity map.
length.\end{abstract}

 ] 

The union of ultrafast lasers and the optical fiber technology has given birth to one of the most important classes of optical sources, and today mode-locked fiber lasers play a central role in both industry and fundamental research \cite{fermann_review_2013, xu_review_2013}. In addition to their wide-spread use as \emph{sources} of ultrashort pulses, mode-locked fiber lasers have also attracted great interest owing to the complex nonlinear dynamics that take place \emph{within} their cavities. This is because ultrafast lasers can be host to a wide variety of dissipative structures and self-organization effects, allowing the devices to be harnessed as convenient test-beds for the exploration of complex systems far from equilibrium \cite{grelu_dissipative_2012, dudley_instabilities_2014}.

Amongst all the dissipative phenomena that can be manifest in mode-locked lasers, \emph{soliton explosions} display perhaps the most striking dynamics. In this regime, a quasi-stable pulse circulates in the cavity for a number of roundtrips, but then suddenly experiences an abrupt structural collapse. Remarkably, within a few roundtrips the collapsed pulse returns back to its initial form, maintaining its integrity until another explosion occurs. This phenomenon was originally identified as a new class of chaotic localized solutions of the complex cubic-quintic Ginzburg-Landau equation (CGLE) \cite{deissler_periodic_1994, soto_pulsating_2000}, and a large number of numerical studies have subsequently been reported in this framework \cite{akhmediev_pulsating_2001, latas_control_2010, cartes_noise_2012, chang_2014}.

Notwithstanding the significant theoretical interest, the existence, characteristics and dynamics underlying the explosion events have not been extensively studied experimentally. This is presumably because the explosions correspond to fleeting transients amidst an ultrafast train of pulses; capturing them requires real-time spectral and temporal diagnostics of a megahertz pulse train. So far only one experimental observation has been reported \cite{cundiff_explosion_2002}. In this work Cundiff {\it et al.} spectrally dispersed the output of a solid-state, Kerr-lens mode-locked Ti:Sapphire laser across a 6-element detector array, and measured the corresponding temporally resolved spectrum. Signatures of explosion events were observed even though the detection apparatus was limited to 12~nm spectral resolution and to averaging over approximately 5 consecutive pulses.  The lack of observations in any other laser configuration raises significant questions pertaining to the experimental ubiquity of the phenomenon: do soliton explosions only manifest themselves in the particular anomalous-dispersion, spatially extended oscillator of ref. \cite{cundiff_explosion_2002}? Compounded by the fact that only spectral signatures have been observed so far, it is clear that significant experimental efforts with high-resolution real-time diagnostics are required to fully understand this phenomenon.

In this Letter, we report on the experimental observation of spectral and temporal signatures of soliton explosions in a mode-locked fiber laser. The explosions appear when the laser is operating in a transition zone~\cite{smirnov_three_2012}, between stable mode-locking~\cite{erkintalo_gco_2012} and noise-like emission~\cite{aguergaray_raman_2013, runge_raman_2014}, and we capture them spectrally using the dispersive Fourier transformation (DFT) \cite{goda_review_2013, wetzel_real_2012, runge_coherence_2013}. Specifically, we record, in real-time and with sub-nanometer resolution, the shot-to-shot spectra emitted by the laser, and identify clear signatures of soliton explosions: abrupt spectral collapses in the output pulse train. We also present explicit time-domain signatures of the explosion events, showing them to be associated with abrupt temporal shifts that can be directly detected at the laser output. All our experimental results are in good agreement with realistic numerical simulations.

\begin{figure}[t]
\centering
\includegraphics[width=\columnwidth,clip = true]{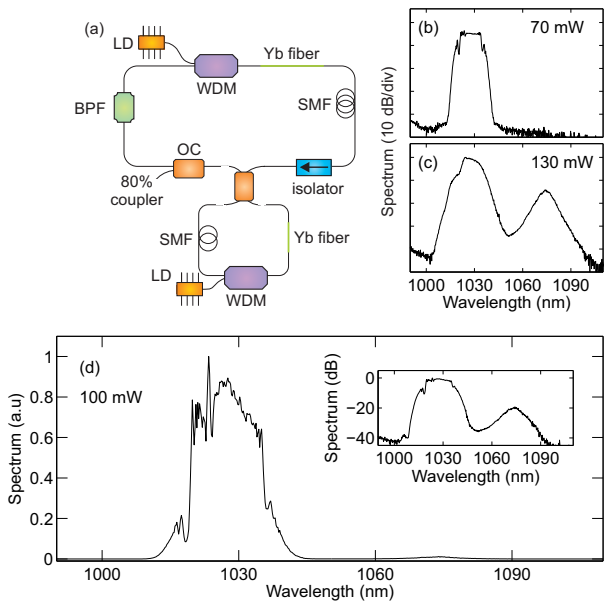}
\vskip-3mm
\caption{\small{(a) Schematic illustration of the laser cavity. LD, laser diode; WDM, wavelength division multiplexer; OC, output coupler; BPF, bandpass filter. (b-d) Average spectra recorded with an OSA when the laser operates in the (b) mode-locking, (c) NL and (d) explosion regime, with pump powers as indicated. Inset in (d) shows the spectrum in logarithmic scale.}}
\label{setup}
\vskip-3mm
\end{figure}

The laser used in our experiment is an all normal-dispersion, all-polarization maintaining passively mode-locked Yb-doped fiber laser \cite{erkintalo_gco_2012, aguergaray_raman_2013, runge_raman_2014}. It is schematically illustrated in Fig.~\ref{setup}(a), and can be seen to be composed of two loops. The main loop includes a segment of Yb-doped fiber followed by a 90~m long segment of polarization maintaining single-mode fiber (SMF). The second loop corresponds to a nonlinear amplifying loop mirror (NALM) which is used as a mode-locker \cite{fermann_nalm_1990}. The NALM contains a short section of Yb-doped fiber and a segment of SMF. An output coupler is located straight after the NALM and it extracts 80~\% of the intracavity power. A narrow bandpass filter (1.7~nm bandwidth) centered at 1028~nm  is used to compensate for the large chirp accumulated over a single roundtrip \cite{chong_andi_2006}. The total cavity length is around 100~m, leading to a fundamental repetition rate of 2~MHz. The two gain sections are pumped by separate but identical 980~nm laser diodes, and mode-locking is achieved by adjusting the pump powers in the two loops. At the laser output we measure the ensemble averaged spectrum using an optical spectrum analyzer (OSA) and use the DFT to acquire high-resolution spectral measurements at the shot-to-shot level. The DFT is implemented by temporally stretching the output of the laser in 10~km of fiber whose group-velocity dispersion is $18.9$~ps$^{2}$km$^{-1}$. This maps the pulse spectra into the temporal domain, allowing us to measure them in real-time using a 12~GHz photodiode and a 12.5~GHz oscilloscope. Overall our DFT configuration allows shot-to-shot measurements with a resolution of approximately 0.2~nm \cite{runge_coherence_2013}.

We have also modelled the laser dynamics using a fully realistic iterative cavity map \cite{runge_andi_2014}. Propagation through all fiber segments, including those in the NALM, is simulated using a generalized nonlinear Schr\"odinger equation that includes both stimulated and spontaneous Raman scattering and higher-order dispersion. The gain dynamics in the active fibers are modelled for given pump powers using the approach in ref. \cite{runge_andi_2014}. Coupling and splice losses are all taken into account.

\begin{figure*}[t]
\centering
\includegraphics[width=\textwidth,clip = true]{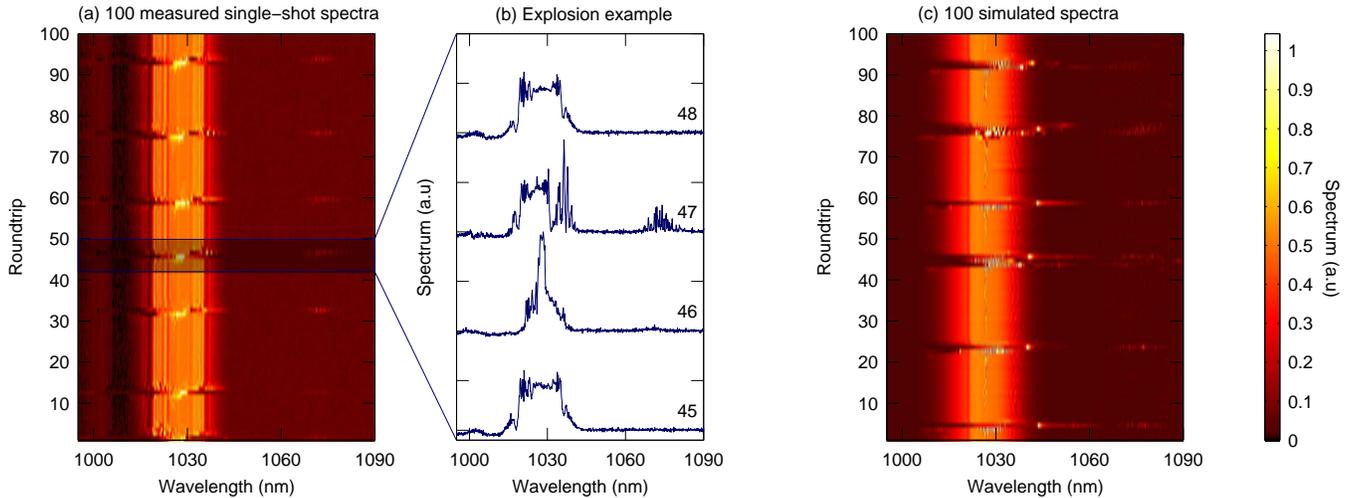}
\caption{\small{(a) Experimentally measured single-shot spectra of 100 consecutive pulses with the laser operating in the transition regime [average spectrum shown in Fig.~\ref{setup}(d)]. Seven soliton explosions can clearly be identified. (b) Example spectra at indicated roundtrip numbers around a particular explosion event. (c) Simulated output spectra for  parameters similar to experiments.}}
\label{single}
\vskip-1mm
\end{figure*}

Depending on the pump levels, this laser configuration can sustain either stable mode-locking or NL pulse emission \cite{aguergaray_raman_2013, runge_raman_2014}. Transition between these two regimes can be achieved by increasing the pump power in the main loop while maintaining constant pump power in the NALM. (The NALM pump power is fixed at 150~mW throughout the paper.) Example averaged spectra for mode-locking and NL regimes are shown in Figs. \ref{setup}(b) and (c), respectively. When the laser is mode-locked, the output spectrum centered at 1028 nm is highly structured, whilst in the NL regime the spectrum is smooth with a strong secondary peak caused by stimulated Raman scattering (SRS) \cite{aguergaray_raman_2013, runge_raman_2014}. By carefully tuning the pump power to lie in between the mode-locking and the NL regimes we observe a third regime, shown in Fig. \ref{setup}(d). At first glimpse, this intermediate mode of operation appears to be a combination of the stable and NL regimes. Indeed, we can still see fine structure reminiscent of the mode-locking regime, but similarly we see a clear Raman component. It is in this transition regime that soliton explosions occur. Accordingly, all the experiments that follow have been performed in this mode of operation.

In order to verify the presence of explosions in the transition regime, we use the DFT to measure the shot-to-shot spectra in real-time. Figure~\ref{single}(a) concatenates 100 experimentally measured single-shot spectra of consecutive pulses emitted by the laser, and we can immediately identify clear signatures~\cite{cundiff_explosion_2002} of soliton explosions. Specifically, when an explosion occurs, the spectrally broad dissipative soliton collapses into a narrower spectrum with higher amplitude, but after a few roundtrips returns back to its previous state until another explosion occurs.  These features are illustrated in more detail in Fig.~\ref{single}(b), where we explicitly show four consecutive spectra around a particular explosion event. In addition to the collapse and revival of the pulse at 1028 nm, we can also see how the explosion events trigger the emission of Raman components at 1075~nm. No such emission is observed in the quasi-stable region; it is the condensation of energy into a narrow spectral band that allows the explosion to act as an efficient Raman pump.

Within the 100 roundtrips shown in Fig.~\ref{single}(a), we can identify 7 clear explosion events, each displaying qualitatively similar characteristics. The events occur intermittently, without any clear periodicity. Explosions can also be observed in our numerical simulations, as shown in Fig.~\ref{single}(c). Here we plot 100 consecutive spectra obtained from simulations using parameters  similar to those in our experiment, and we observe good qualitative agreement with measured results.

\begin{figure}[htb]
\centering
\includegraphics[width=\columnwidth,clip = true]{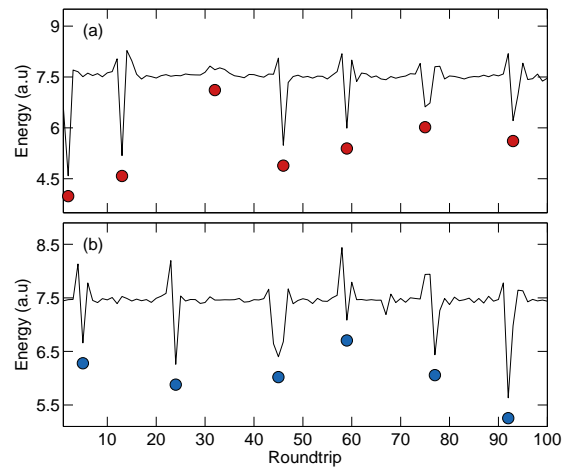}
\vskip-3mm
\caption{\small{Energies of 100 consecutive output pulses extracted from (a) experimental measurements and (b) numerical simulations. Corresponding single-shot spectra are shown in Fig.~\ref{single}. The solid circles mark the explosions.}}
\label{energy}
\vskip-3mm
\end{figure}

In order to gain more insights we investigated the characteristics and dynamics of the explosion events. We first studied the evolution of the energy per pulse
by integrating the measured and simulated spectra over the complete spectral band. Experimental and numerical results are shown in Fig.~\ref{energy}(a) and (b), respectively, and the two can again be seen to be in good agreement. (The solid circles mark the roundtrips at which the explosions occur.) We can see that the energy increases just before an explosion, but then suddenly drops before bouncing back to its previous level. This should be contrasted with previous studies that have exclusively predicted explosions to be associated with an increase in total energy \cite{akhmediev_pulsating_2001, cundiff_explosion_2002}. However, we can understand the decrease observed in our configuration by recalling that during an explosion significant part of the total energy is converted to a Raman pulse through SRS. This wave is temporally dispersed in the cavity and will therefore be attenuated by the NALM~\cite{runge_raman_2014}, yielding an overall decrease in total energy.

\begin{figure}[t]
\centering
\includegraphics[width=\columnwidth,clip = true]{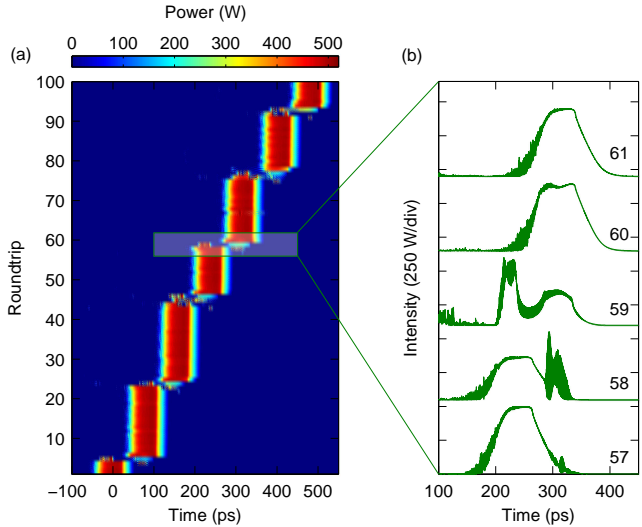}
\vskip -3mm
\caption{\small{(a) Temporal evolution of the simulated output pulse envelope over 100 consecutive roundtrips. (b) Zoom on 5 consecutive roundtrips as indicated, showing the pulse shifting from one pulse to another during an explosion event.}}
\label{time_simu}
\vskip -3mm
\end{figure}

We next investigated the time-domain dynamics associated with explosion events. To this end, Fig.~\ref{time_simu}(a) shows the roundtrip-to-roundtrip evolution of the temporal pulse envelope corresponding to the simulation in Fig. \ref{single}(c). Note that the simulation is performed in a reference frame that moves at the group velocity of the stable pulses, which therefore appear stationary in the evolution plot. In contrast, roundtrips showing spectral explosions can be seen to be associated with abrupt temporal shifts. Similar shifts have previously been reported in CGLE-based theoretical studies, and they have been linked to the presence of higher-order effects \cite{latas_control_2010}. Our simulations show that, in our configuration, it is the strong SRS perturbation that gives rise to the extreme timing jitter. Interestingly, closer investigation of the temporal envelope [see Fig.~\ref{time_simu}(b)] shows that the explosion resembles the onset of double-pulsing \cite{{bale_transition_2009}}. Indeed, a secondary pulse can be clearly seen to develop at the trailing edge of the first pulse. The two pulses compete but ultimately cannot coexist; the first pulse vanishes and only the second, trailing pulse remains. It is the switch from the initial to the final pulse that gives rise to the sudden temporal shift.

\begin{figure}[!t]
\centering
\includegraphics[width=\columnwidth,clip = true]{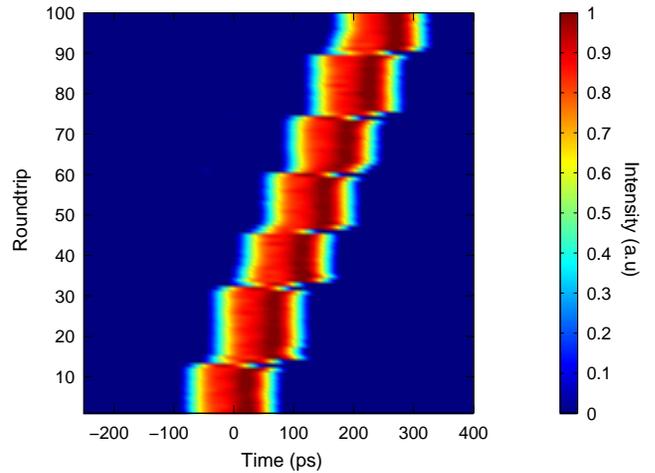}
\vskip -3mm
\caption{\small{Experimentally measured temporal evolution relative to the average roundtrip time over 100 roundtrips.}}
\label{shift}
\vskip -3mm
\end{figure}

In our simulations each explosion is associated with a time shift of approximately 90~ps. This is sufficiently large to be discerned with a fast oscilloscope, implying the possibility of observing the explosion events directly in the time-domain. We tested this prediction by recording 100 consecutive pulses immediately at the laser output when the laser was operating in the explosion regime. We looked for abrupt temporal shifts by dividing the recorded real-time signal into segments whose duration equals the average cavity roundtrip time. We then concatenated all the segments into a single false color plot. The results are shown in Fig. \ref{shift}, and we can clearly identify abrupt jumps in the time-domain signal. During each jump, the pulse exits the cavity approximately 40~ps later than expected based on the average roundtrip time, which is in reasonable qualitative agreement with numerical observations. Note that this  timing jitter does not impair our spectral DFT measurements: The 40 ps shift maps into a negligible 0.1~nm wavelength jitter. It should also be stressed that we do not observe similar temporal jumps when the laser operates in the fully stable mode-locking regime. These measurements therefore confirm that soliton explosions can give rise to extreme temporal shifts, enabling the events to be directly observed in the time-domain.

We close by emphasizing that the explosions are not unique to the specific 100 m long cavity configuration discussed above. Indeed, by controlling the amount of SMF in the device we have performed additional experiments for a wide range of cavity lengths, and observed explosions for all realizations whose lengths lie between 90 to 200~m. We find that the explosions always appear in the transition zone between the stable and the noise-like regimes. When the cavity is shorter (longer) than 90 m (200 m), the laser only sustains the stable (NL) regime. Accordingly, explosions are not observed since there is no transition regime that could host them. These observations are in good agreement with previous analyses~\cite{akhmediev_pulsating_2001,cundiff_explosion_2002}, where explosions were found to exist near the threshold of unstable operation. Finally, whilst the qualitative characteristics of the explosion events always display characteristics similar to those described above, we have found that the frequency of their occurrence increases with the cavity length and the pump power. This is again in good agreement with experiments in a Ti:sapphire oscillator, where the cavity dispersion and pump power was found to influence the explosion frequency~\cite{cundiff_explosion_2002}.

In conclusion, we have reported the first experimental observation of soliton explosions in a fiber laser. We have recorded the roundtrip-to-roundtrip spectra of a passively mode-locked fiber laser, identifying clear explosion signatures when operating in a regime between stable and noise-like emission. Our observations are the first of their kind in a normal dispersion oscillator, and we have also reported the first direct time-domain signatures of explosion events. All these results are in good agreement with realistic lumped cavity simulations. Our work demonstrates that soliton explosions can manifest themselves between two ubiquitous operation regimes of mode-locked fiber lasers, and that they can be experimentally investigated using a simple yet high-resolution technique. We therefore expect these results to pave way for extensive experimental investigations, allowing the dynamics and characteristics of soliton explosions and related dissipative structures to be fully unveiled.

\end{document}